\documentclass
[superscriptaddress,secnumarabic,amssymb,amsmath,nobibnotes,aps,prd,showkeys,showpacs,nofootinbib,onecolumn,12pt]{revtex4}%
\usepackage{graphicx}
\usepackage{epsf}
\usepackage{bm}
\usepackage{amsmath}
\usepackage{amsfonts}
\usepackage{amssymb}%
\providecommand{\U}[1]{\protect\rule{.1in}{.1in}}

\newcommand{\be}{\begin{equation}}
\newcommand{\ee}{\end{equation}}

\newcommand{\mincir}{\raise
-3.truept\hbox{\rlap{\hbox{$\sim$}}\raise4.truept\hbox{$<$}\ }}
\newcommand{\magcir}{\raise
-3.truept\hbox{\rlap{\hbox{$\sim$}}\raise4.truept\hbox{$>$}\ }}

\begin{document}
\title{Anisotropic Spacetimes in Chiral Scalar Field Cosmology}
\author{A. Giacomini}
\email{alexgiacomini@uach.cl}
\affiliation{Instituto de Ciencias F\'{\i}sicas y Matem\'{a}ticas, Universidad Austral de
Chile, Valdivia 5090000, Chile}
\author{P.G.L. Leach}
\email{leachp@ukzn.ac.za}
\affiliation{Institute of Systems Science, Durban University of Technology, Durban 4000,
South Africa}
\author{G. Leon}
\email{genly.leon@ucn.cl}
\affiliation{Departamento de Matem\'{a}ticas, Universidad Cat\'{o}lica del Norte, Avda.
Angamos 0610, Casilla 1280 Antofagasta, Chile}
\author{A. Paliathanasis}
\email{anpaliat@phys.uoa.gr}
\affiliation{Instituto de Ciencias F\'{\i}sicas y Matem\'{a}ticas, Universidad Austral de
Chile, Valdivia 5090000, Chile}
\affiliation{Institute of Systems Science, Durban University of Technology, Durban 4000,
South Africa}

\begin{abstract}
We study the behaviour and the evolution of the cosmological field equations
in an homogeneous and anisotropic spacetime with two scalar fields coupled in
the kinetic term. Specifically, the kinetic energy for the scalar field
Lagrangian is that of the Chiral model and defines a two-dimensional maximally
symmetric space with negative curvature. For the background space we assume
the locally rotational spacetime which describes the Bianchi I, the Bianchi
III and the Kantowski-Sachs anisotropic spaces. We work on the $H$%
-normalization and we investigate the stationary points and their stability.
For the exponential potential we find a new exact solution which describes an
anisotropic inflationary solution. The anisotropic inflation is always
unstable, while future attractors are the scaling inflationary solution or the
hyperbolic inflation. For scalar field potential different from the
exponential, the de Sitter universe exists.

\end{abstract}
\keywords{Multifield Cosmology; Chiral Cosmology; Dynamical analysis; Kantowski-Sachs}
\pacs{98.80.-k, 95.35.+d, 95.36.+x}
\date{\today}
\maketitle

\section{Introduction}

\label{sec1}

Gravitational models with two or more scalar fields for the description of the
matter part for the gravitational field equations have been widely studied in
the literature during recent years
\cite{mf0,mf1,mf2,mf3,mf4,mf5,mf6,mf7,mf8,mf9,mf10,mf11,mf12}. Multi-scalar
field cosmological models have been used as alternative mechanisms for the
description of inflation \cite{guth} as also as unified dark energy models.
Indeed, because of the additional degrees of freedom provided by the scalar
field, the exit from the inflationary era is different from the single-scalar
field theory. Specifically, it is not necessary the values for the scalar
fields to be the same at the beginning of the inflation and at the end of the
inflation. Hence, the curvature perturbations can be affected by the different
number of e-folds \cite{g1,g2}. On the other hand, multi-scalar fields provide
non-adiabatic field perturbations which generate observable non-Gaussianities
in the power spectrum \cite{g3,g4,g5}. As far as the late time universe is
concerned, multi-scalar field models provide dark energy models which can
cross the phantom divide line without the presence of ghosts \cite{qq2}, as
also to describe the dark matter component of the universe \cite{ancqg}.

In this study we focus upon the asymptotic dynamics for the field equations in
a two-scalar field theory known as the Chiral model within an homogeneous and
anisotropic background space \cite{atr6,vr91,vr92,ts1}. The kinetic energy of
the two scalar field lies on a two-dimensional maximally symmetric space of
negative curvature, hyperbolic space. This multi-scalar field model provides
the so-called hyperbolic inflation \cite{vr91,s1}. However, there are various
applications of this model and in other areas of the cosmic evolution
\cite{ts1,ts2,ts3,ts4,ts5,ts6,ts7,ts8,ts9,ts10}.

On the other hand, homogeneous and anisotropic are mainly expressed by the
Bianchi class of spatially homogeneous spacetimes. Bianchi spacetimes have
been mainly used for the discussion of anisotropies in the very early universe
\cite{Mis69,jacobs2,collins}. The presence of a cosmological constant in
Bianchi spacetimes leads to isotropic universe as a future
solution~\cite{wald}. For the physical space in this study we assume the
generic line element \cite{nilson}
\begin{equation}
ds^{2}=-dt^{2}+e^{2\alpha\left(  t\right)  }\left(  e^{2\beta\left(  t\right)
}dx^{2}+e^{-\beta\left(  t\right)  }\left(  dy^{2}+f^{2}\left(  y\right)
dz^{2}\right)  \right)  , \label{mm.01}%
\end{equation}
where the function $f\left(  y\right)  $ has one of the following forms,
$f_{A}\left(  y\right)  =1$, $f_{B}\left(  y\right)  =\sinh\left(
\sqrt{\left\vert K\right\vert }y\right)  $ and $f_{C}\left(  y\right)
=\sin\left(  \sqrt{\left\vert K\right\vert }y\right)  $. The line element
(\ref{mm.01}) corresponds to homogeneous locally rotational spacetimes (LRS)
induced with four isometries. For $f_{A}\left(  y\right)  $ the spacetime is
that of Bianchi I, for $f_{B}\left(  y\right)  $ is that of the Bianchi III
metric while for $f_{C}\left(  y\right)  $ the line element reduces to that of
Kantowski-Sachs.\ These three different families of spacetimes reduce to the
flat, closed and open Friedmann--Lema\^{\i}tre--Robertson--Walker (FLRW)
spacetimes when the parameter, $\beta\left(  t\right)  $, becomes constant.
Indeed, the parameter, $\beta\left(  t\right)  $, is the anisotropic parameter
while $\alpha\left(  t\right)  $ is the scale factor for the three-dimensional
hypersurface. These spacetimes play an important role on the description of
the very early universe and specifically during the pre-inflationary era
\cite{kas2,kas3,kas4,kas5,kas6,kas7,kas8,kas9,kas10}.

In the following we investigate the asymptotic dynamics and the evolution for
the field equations by investigate the stationary points for the field
equations \cite{aco}. Every stationary point describes a specific era for the
evolution of the field equations \cite{sc1}. The analysis of the stability
properties for the stationary points is essential in order to construct the
cosmological history \cite{sc2,sc3}. Such an analysis provides important
information for the viability of a given gravitational theory \cite{sc4}. In
addition this analysis provides important information about the initial
condition problem. Such analysis has been widely studied in various
gravitational models \cite{sc5,sc6,sc7} while some studies in anisotropic
universes can be found in \cite{sc8,sc9,sc10,sc11,sc12,sc13,sc14}. The plan of
the paper is as follows.

In Section \ref{sec2} we present the field equations for the Chiral theory
with anisotropic background space described by the anisotropic line element
(\ref{mm.01}). Section \ref{sec3} includes the new results of this analysis in
which we study the general evolution and the asymptotic behaviour for the
field equations for the Chiral theory for the potential function of the
hyperbolic inflation \cite{vr91}. In Section \ref{sec4} we investigate the
dynamics for a scalar field potential beyond the exponential. Finally, in
Section \ref{sec5} we summarize the results and we draw our conclusions.

\section{Field equations}

\label{sec2}

We assume the gravitational model in a Riemann manifold $g_{\mu\nu}\left(
x^{\kappa}\right)  $ and Ricci scalar $R\left(  x^{\kappa}\right)  ~$with
two-scalar fields minimally coupled to the gravity, which is described by the
following Action Integral.%
\begin{equation}
S=\int\sqrt{-g}dx^{4}\left(  \frac{R}{2}+L_{C}\left(  \phi,\nabla_{\mu}%
\phi,\psi,\nabla_{\mu}\psi\right)  \right)  . \label{ch.01}%
\end{equation}

Lagrangian $L_{C}\left(  \phi,\nabla_{\mu}\phi,\psi,\nabla_{\mu}\psi\right)  $
is assumed to be that of the Chiral model, that is%
\begin{equation}
L_{C}\left(  \phi,\nabla_{\mu}\phi,\psi,\nabla_{\mu}\psi\right)  =-\frac{1}%
{2}g^{\mu\nu}\left(  \nabla_{\mu}\phi\nabla_{\nu}\phi+e^{-2\kappa\phi}%
\nabla_{\mu}\psi\nabla_{\nu}\psi\right)  +V(\phi). \label{ch.02}%
\end{equation}

Consequently, the two scalar fields, $\phi$ and $\psi$, are defined on a
two-dimensional space of constant and negative curvature, while their
evolution is defined on the physical space with metric $g_{\mu\nu}.$ In the
following we assume that $\kappa\neq0$ and the scalar field potential$~$is
that of the hyperbolic inflation, that is, $V\left(  \phi\right)
=V_{0}e^{-\lambda\phi}$.

For the line element (\ref{mm.01}) we derive%
\begin{equation}
R\left(  \alpha,\dot{\alpha},\beta,\dot{\beta}\right)  =6\ddot{\alpha}%
+12\dot{\alpha}^{2}+\frac{3}{2}\dot{\beta}^{2}-2e^{\beta-2\alpha}K
\end{equation}
and $\sqrt{-g}=e^{3\alpha}$, where the overdot means total derivative with
respect the independent parameter $t$, i.e. $\dot{\alpha}=\frac{d\alpha}{dt}$.

Hence, by substituting into (\ref{ch.01}) and assuming that the scalar fields
inherit the symmetries of the background space, we obtain the following system
of second-order differential equations \cite{anuni1}
\begin{equation}
2\ddot{\alpha}+3\dot{\alpha}^{2}+\frac{3}{4}\dot{\beta}^{2}+\frac{1}{2}\left(
\dot{\phi}^{2}+e^{-2\kappa\phi}\dot{\psi}\right)  -V\left(  \phi\right)
-\frac{1}{3}e^{-2\alpha-\beta}K=0, \label{ch.05}%
\end{equation}%
\begin{equation}
\ddot{\beta}+3\dot{\alpha}\dot{\beta}+\frac{2}{3}e^{-2\alpha-\beta}K=0,
\label{ch.06}%
\end{equation}%
\begin{equation}
\ddot{\phi}+\kappa e^{-2\kappa\phi}\dot{\psi}^{2}+3\dot{\alpha}\dot{\phi
}+V_{,\phi}=0, \label{ch.07}%
\end{equation}%
\begin{equation}
\ddot{\psi}-2\kappa\dot{\phi}\dot{\psi}+3\dot{\alpha}\dot{\psi}=0
\label{ch.08}%
\end{equation}
and the constraint equation
\begin{equation}
e^{3\alpha}\left(  3\dot{\alpha}^{2}-\frac{3}{4}\dot{\beta}^{2}-\frac{1}%
{2}\left(  \dot{\phi}^{2}+e^{-2\kappa\phi}\dot{\psi}^{2}\right)  -V\left(
\phi\right)  \right)  -e^{\alpha-\beta}K=0. \label{ch.04}%
\end{equation}

The parameter $K$ denotes the spatial curvature of the three-dimensional
hypersurface part for (\ref{mm.01}). Indeed, for Bianchi I space $K=0$, for
Bianchi III space is positive $K>0$ while for the Kantowski-Sachs space, $K<0$.

\subsection{Dimensionless variables}

In order to study the global evolution of the field equations we define the
new set of variables
\begin{equation}
\Sigma=\frac{\dot{\beta}}{2H}~,~x=\frac{\dot{\phi}}{\sqrt{6}H}~,~y^{2}%
=\frac{V\left(  \phi\right)  }{3H^{2}}~, \label{ch.10}%
\end{equation}%
\begin{equation}
z=\frac{e^{-\kappa\phi}\dot{\psi}}{\sqrt{6}H}~,~\omega_{R}=\frac{R^{\left(
3\right)  }}{3H^{2}}~, \label{ch.11}%
\end{equation}
where $R^{\left(  3\right)  }=e^{\alpha-\beta}K$ and $H\left(  t\right)
=\dot{\alpha}$ is the expansion rate.

Moreover, we select the new independent variable to be $dt=d\tau~,~\tau
=\alpha$. Thus, in the new variables the field equations (\ref{ch.05}%
)-(\ref{ch.08}) read%
\begin{equation}
\Sigma^{\prime}=-y^{2}\left(  1+\Sigma\right)  +\left(  2\Sigma-1\right)
\left(  x^{2}+z^{2}+\Sigma^{2}-1\right)  ~, \label{ch.21}%
\end{equation}%
\begin{equation}
x^{\prime}=2x^{3}+\frac{\sqrt{6}}{2}\left(  y^{2}\lambda-2z^{2}\kappa\right)
-x\left(  y^{2}-2\left(  z^{2}+\Sigma^{2}-1\right)  \right)  , \label{ch.22}%
\end{equation}%
\begin{equation}
y^{\prime}=\frac{1}{2}y\left(  2\left(  1-y^{2}\right)  +4\left(  x^{2}%
+z^{2}+\Sigma^{2}\right)  -\sqrt{6}x\lambda\right)  \label{ch.23}%
\end{equation}
and
\begin{equation}
z^{\prime}=z\left(  \sqrt{6}x\kappa+2\left(  x^{2}+z^{2}+\Sigma^{2}-1\right)
-y^{2}\right)  ~, \label{ch.24}%
\end{equation}
where $V\left(  \phi\right)  =V_{0}e^{-\lambda\phi}$, $\lambda=\frac{V_{,\phi
}}{V}$ and $\Sigma^{\prime}=\frac{d\Sigma}{d\tau}$. Furthermore, the
constraint equation (\ref{ch.04}) reduces to the following algebraic equation%
\begin{equation}
\omega_{R}=1-\left(  \Sigma^{2}+x^{2}+y^{2}+z^{2}\right)  . \label{ch.26}%
\end{equation}

By definition, the parameter $y$ is positive, while the field equations remain
invariant under the discrete transformation $z\rightarrow-z$. Hence we select
to work with $z>0$.

Moreover, we define the deceleration parameter $q=-1-\frac{\ddot{a}}{\dot
{a}^{2}}$, which with the use of the dimensionless variables is
\begin{equation}
q\left(  \Sigma,x,y,z\right)  =2\left(  x^{2}+z^{2}+\Sigma^{2}\right)
-y^{2}\text{.} \label{ch.27}%
\end{equation}

\section{Asymptotic dynamics}

\label{sec3}

We continue our analysis with the study of the dynamics provided by the
dynamical system (\ref{ch.21})-(\ref{ch.26}). Specifically, we determine the
stationary points and we investigate their stability. Every stationary point
corresponds to a specific era in the evolution of the cosmological history.

We summarize the stationary points, $P=\left(  \Sigma\left(  P\right)
,x\left(  P\right)  ,y\left(  P\right)  ,z\left(  P\right)  \right)  $, in
three categories, (A) stationary points of General Relativity; (B) stationary
points of quintessence and (C) stationary points with two scalar fields. The
first family of stationary points describes exact solutions without any matter
source. Thus, the exact solutions described by these points are these of
General Relativity in the vacuum~$\left(  x,y,z\right)  =\left(  0,0,0\right)
$. For the family (B) of points, only the scalar field $\phi$ contributes in
the cosmological solution, that is~$z=0,~\left(  x,y\right)  \neq\left(
0,0\right)  $, while for the third family of points, both the scalar fields
contribute, i.e. $\dot{\phi}\dot{\psi}\neq0$. It is important to mention that
the stability properties of the points on the families (A) and (B) depend upon
the existence of the second field, that is, of the dynamical variable $z$.
Thus we should perform a detailed analysis on the stability conditions.
Moreover, stationary points with $\Sigma=0$, correspond to isotopic background
space, while stationary points with $\eta=0$, indicate that the exact solution
is a static solution. Furthermore, the background space in a asymptotic
solution is that of Bianchi I or spatially flat FLRW metric when $\omega
_{R}=0,$ of Bianchi III or closed FLRW metric when $\omega_{R}>0$, or
Kantowski-Sachs or open FLRW universe when $\omega_{R}<0$.

We determine the stationary points for the dynamical system (\ref{ch.21}%
)-(\ref{ch.26}) for values of the dynamical variables in the finite regime.

\subsection{Stationary points of family A}

The stationary points which belong to the family A are%
\[
A_{1}^{\pm}=\left(  \pm1,0,0,0\right)  ~,~A_{2}=\left(  \frac{1}%
{2},0,0,0\right)  .
\]
For each of the stationary points we calculate $\omega_{R}\left(  A_{1}^{\pm
}\right)  =0~$ and $\omega_{R}\left(  A_{2}^{\pm}\right)  =\frac{3}{4}$.~

We continue with the discussion of the physical properties for the asymptotic
solutions at the stationary points while we investigate the stability
properties of the points.

Points $A_{1}^{\pm}$ describe anisotropic spacetime with zero spatial
curvature, that is, the asymptotic solution at the points correspond to Kasner
universes. The eigenvalues of the linearized system near the asymptotic points
are $e_{1}\left(  A_{1}^{+}\right)  =6~,~e_{2}\left(  A_{1}^{+}\right)  =3$,
$e_{3}\left(  A_{1}^{+}\right)  =0$ and $e_{4}\left(  A_{1}^{+}\right)  =0~$;
$e_{1}\left(  A_{1}^{-}\right)  =3~,~e_{2}\left(  A_{1}^{-}\right)  =2$,
$e_{3}\left(  A_{1}^{-}\right)  =0$ and $e_{4}\left(  A_{1}^{-}\right)  =0$.
Thus, points $A_{1}^{\pm}$ are sources and the asymptotic solutions are always unstable.

Point $A_{2}$ describes an anisotropic exact solution with nonzero spatial
curvature. The eigenvalues are $e_{1}\left(  A_{3}^{\pm}\right)  =\frac{3}%
{2}~,~e_{2}\left(  A_{3}^{\pm}\right)  =-\frac{3}{2}$, $e_{3}\left(
A_{3}^{\pm}\right)  =-\frac{3}{2}$ and $e_{4}\left(  A_{3}^{\pm}\right)
=-\frac{3}{2}$~ from which we conclude that the vacuum anisotropic solution is
unstable, while point $A_{2}$ is a saddle point for the dynamical system

For all the stationary points we derive positive value for the deceleration
parameter, $q\left(  A_{1}^{\pm}\right)  =2$,~$q\left(  A_{2}\right)
=\frac{1}{2}$.

\subsection{Stationary points of family B}

The family B for the stationary points of the dynamical system (\ref{ch.21}%
)-(\ref{ch.26}) is consists of the points,%
\[
B_{1}^{\pm}=\left(  \pm\sqrt{1-x^{2}},x,0,0\right)  ~,~B_{2}=\left(
0,\frac{\lambda}{\sqrt{6}},\sqrt{1-\frac{\lambda^{2}}{6}},0\right)
\]%
\[
B_{3}=\left(  \frac{1}{2}\left(  1-\frac{3}{2\left(  1+\lambda^{2}\right)
}\right)  ,\frac{\sqrt{6}\lambda}{2\left(  1+\lambda^{2}\right)  },\frac
{\sqrt{6\left(  2+\lambda^{2}\right)  }}{2\left(  1+\lambda^{2}\right)
},0\right)
\]
with $\omega_{R}\left(  B_{1}^{\pm}\right)  =0$,~$\omega_{R}\left(  B_{2}%
^{\pm}\right)  =0$, and $\omega_{R}\left(  B_{3}^{\pm}\right)  =\frac{3}%
{4}\frac{\left(  \lambda^{2}-4\right)  }{\left(  1+\lambda^{2}\right)  ^{2}}$.

\ Points $B_{1}^{\pm},$ exist when $x^{2}\leq1$, for $x^{2}<1$. They describe
families of points where the asymptotic solutions are Kasner universes,
respectively. Moreover, for $x^{2}=1$, the solution is that of spatially flat
FLRW universe dominated by a stiff fluid. For the asymptotic solutions at the
stationary points the decelerating parameter is derived to be $q=2$. Hence,
the points do not describe acceleration. The eigenvalues are derived to be
$e_{1}\left(  B_{1}^{\pm}\right)  =2\left(  2\mp\sqrt{1-x^{2}}\right)
$,~$e_{2}\left(  B_{1}^{\pm}\right)  =\sqrt{6}\kappa x$~, $e_{3}\left(
B_{1}^{\pm}\right)  =\frac{1}{2}\left(  6-\sqrt{6}\kappa\lambda\right)  $
and$~e_{4}\left(  B_{1}^{\pm}\right)  =0$. Consequently, points $B_{1}^{\pm}$
are sources or saddle points. Specifically, points $B_{2}^{+}$ are sources
when $\left\{  \kappa<0,-1\leq x<0,\lambda>\frac{\sqrt{6}}{x}\right\}  $ or
$\left\{  \kappa>0,0<x\leq1,\lambda<\frac{\sqrt{6}}{x}\right\}  $. Otherwise
they are saddle points.

\begin{figure}[ptb]
\centering\includegraphics[width=1\textwidth]{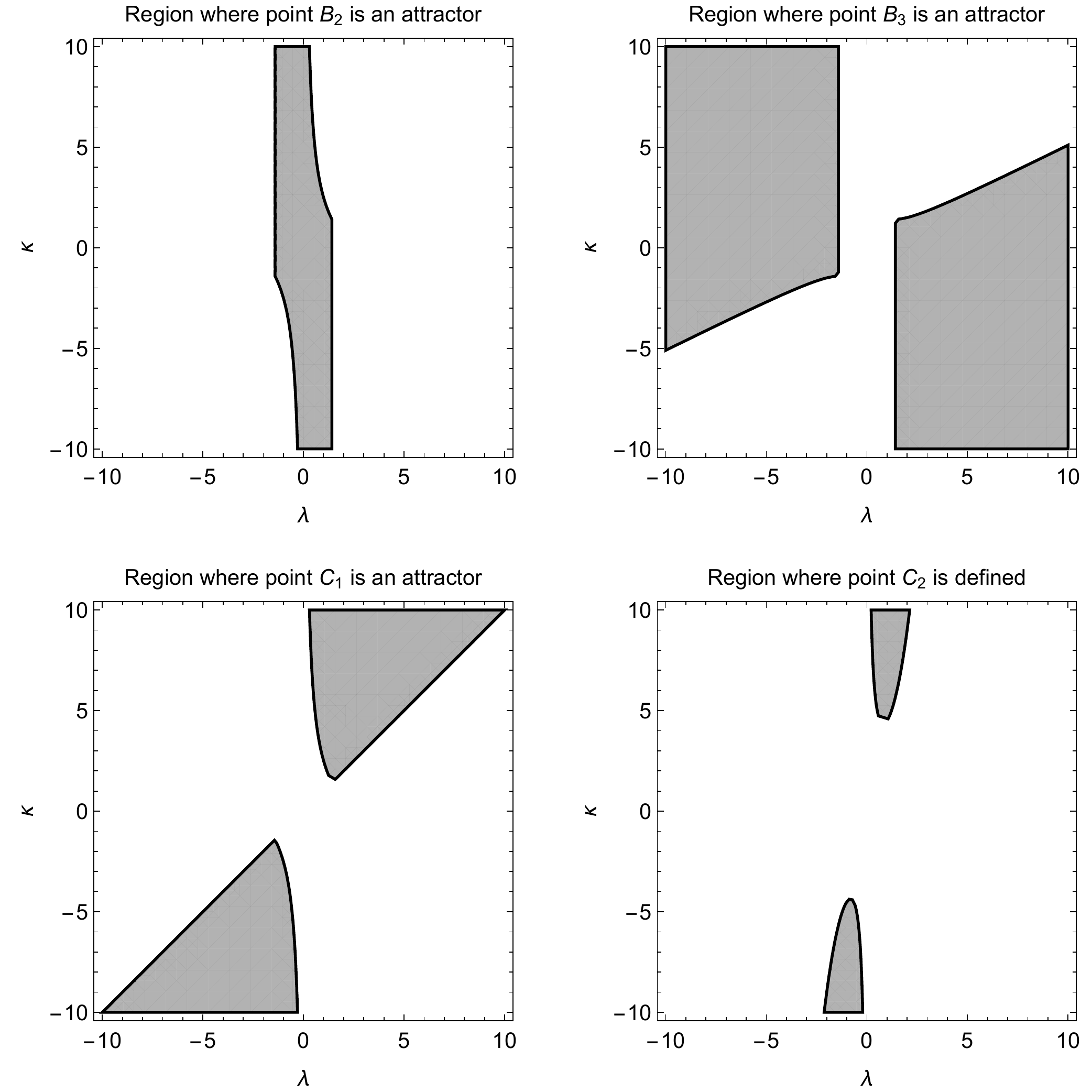}\caption{Region plot in
the space of the free parameters $\left(  \lambda,\kappa\right)  $ where the
asymptotic solutions at points $B_{2}$,~$B_{3}$ and $C_{1}$ are stable
solutions and points are attractors, and when point $C_{2}$ is physical
accepted and describes anisotropic inflation. }%
\label{fig0}%
\end{figure}

Point $B_{2}$ is real and physical accepted when $\lambda^{2}\leq6$ and
describes \ the scaling solution for the quintessence field with the
exponential scalar field potential in a spatially flat FLRW background space.
The deceleration parameter is calculated to be $q\left(  B_{2}\right)
=\frac{\lambda^{2}-2}{2}$, which means that for $\lambda^{2}<2$ the asymptotic
solution describes an accelerating universe. The eigenvalues of the linearized
system are $e_{1}\left(  B_{2}\right)  =\frac{\lambda^{2}+2\kappa\lambda-6}%
{2}~,~e_{2}\left(  B_{2}\right)  =\frac{\lambda^{2}-6}{2}$ ,$~e_{3}\left(
B_{2}\right)  =\frac{\lambda^{2}-6}{2}$,$~e_{4}\left(  B_{2}\right)  =\left(
\lambda^{2}-2\right)  $. Hence, the asymptotic solution at the point $B_{2}$
is stable and the point $B_{2}$ is an attractor when $\left\{  -\sqrt
{2}<\lambda<0,\kappa>\frac{6-\lambda^{2}}{2\lambda}\right\}  $ or $\left\{
0<\lambda<\sqrt{2},\kappa<\frac{6-\lambda^{2}}{2\lambda}\right\}  $. \ The
region where point $B_{2}$ is an attractor is presented in Fig. \ref{fig0}.

Point $B_{3}$ describes the exact solution with anisotropic spacetime, \qquad
with positive spatial curvature when $\lambda^{2}>4$, or with negative spatial
curvature when for $\lambda^{2}<4$, while for $\lambda^{2}=4$ it describes a
Bianchi I universe. The deceleration parameter is $q\left(  B_{3}\right)
=\frac{\lambda^{2}-2}{2\left(  \lambda^{2}+1\right)  }$. Hence there is
acceleration for $\lambda^{2}<2$. The eigenvalues of the linearized system are
~$e_{1}\left(  B_{3}\right)  =-\frac{3\left(  2-2\kappa\lambda+\lambda
^{2}\right)  }{2\left(  1+\lambda^{2}\right)  }$ , $e_{2}\left(  B_{3}\right)
=-\frac{3\left(  2+3\lambda^{2}+\lambda^{4}\right)  }{2\left(  1+\lambda
^{2}\right)  }$ ,~$e_{3}\left(  B_{3}\right)  =-\frac{3\left(  2+3\lambda
^{2}+\lambda^{4}+\sqrt{\left(  1+\lambda\right)  ^{2}\left(  2+\lambda
^{2}\right)  \left(  7\lambda^{2}-18\right)  }\right)  }{4\left(
1+\lambda^{2}\right)  ^{2}}$ and $e_{4}\left(  B_{3}\right)  =-\frac{3\left(
2+3\lambda^{2}+\lambda^{4}\pm\sqrt{\left(  1+\lambda\right)  ^{2}\left(
2+\lambda^{2}\right)  \left(  7\lambda^{2}-18\right)  }\right)  }{4\left(
1+\lambda^{2}\right)  ^{2}}$. In Fig. \ref{fig0} we present the region in the
space $\left(  \lambda,\kappa\right)  $ in which the point $B_{3}$ is an
attractor. \ 

\subsection{Stationary points of family C}

The third family of stationary points for the dynamical (\ref{ch.21}%
)-(\ref{ch.26}) includes the points
\[
C_{1}=\left(  0,\frac{\sqrt{6}}{2\kappa+\lambda},\sqrt{\frac{2\kappa}%
{2\kappa+\lambda}},\frac{\sqrt{\lambda^{2}+2\kappa\lambda-6}}{2\kappa+\lambda
}\right)  ~,
\]%
\[
C_{2}=\left(  2-\frac{6\kappa}{2\kappa+\lambda},\frac{\sqrt{6}}{2\kappa
+\lambda},\frac{\sqrt{6\kappa\left(  2\kappa-\lambda\right)  }}{2\kappa
+\lambda},\frac{\sqrt{6\kappa\lambda-3\left(  \lambda^{2}+2\right)  }}%
{2\kappa+\lambda}\right)  ~\text{.}%
\]

The stationary point, $C_{1}$, exists when $2\kappa+\lambda\neq0$ and
$\left\{  \lambda\leq-\sqrt{6},\kappa<0\right\}  $, $\left\{  -\sqrt
{6}<\lambda<0,\kappa<\frac{6-\lambda^{2}}{2\lambda}\right\}  ~$,$~\left\{
0<\lambda<\sqrt{6},\kappa>\frac{6-\lambda^{2}}{2\lambda}\right\}
~$,$~\left\{  \lambda\geq\sqrt{6},\kappa>0\right\}  $ ,~while the spatial
curvature for the background space is zero, that is, $\omega_{R}\left(
C_{1}\right)  =0$. Hence the background space at the stationary point is that
of spatially flat\ FLRW universe. The point $C_{1}$ describes hyperbolic
inflation \cite{vr91}. The deceleration parameter is written as $q\left(
C_{1}\right)  =2-\frac{6\kappa}{2\kappa+\lambda}$. Hence, the asymptotic
solution at $C_{1}$ describes inflationary isotropic inflation when $\left\{
\lambda\leq-\sqrt{2},\kappa<\lambda\right\}  $, $\left\{  -\sqrt{2}%
<\lambda<0,\kappa<\frac{6-\lambda^{2}}{2\lambda}\right\}  ~$,$~\left\{
0<\lambda<\sqrt{2},\kappa>\frac{6-\lambda^{2}}{2\lambda}\right\}  ~$ and
$~\left\{  \lambda\geq\sqrt{2},\kappa>\lambda\right\}  $. In Fig. \ref{fig0}
we present the region in the space $\left(  \lambda,\kappa\right)  $ in which
the point $C_{1}$ is an attractor. Note that, when $C_{1}$ is an attractor,
the asymptotic solution describes an inflationary universe, i.e. $q\left(
C_{1}\right)  <0$.

The point $C_{2}$ describes anisotropic solutions with $\omega_{R}\left(
C_{2}\right)  =-\frac{12\kappa\left(  \kappa-\lambda\right)  }{\left(
2\kappa+\lambda\right)  ^{2}}$. The points are real and physical acceptable
when~$2\kappa\lambda>\left(  2+\lambda^{2}\right)  ^{2}.~$ The deceleration
parameter is derived to be $q\left(  C_{2}\right)  =2-\frac{6\kappa}%
{2\kappa+\lambda}.~$Consequently, when the point is real and physical
accepted, it describes accelerated anisotropic inflationary solution in a
Kantowski-Sachs background space. We derive the eigenvalues of the linearized
system and we find that the four eigenvalues do not have real parts with
negative values, for the same values of the parameters, $\lambda$ and $\kappa
$. Moreover, they do not have real parts with positive values for the same
values of the parameters, $\lambda$ and $\kappa$. We conclude that the
asymptotic anisotropic solution is unstable and point $C_{2}$ is a saddle
point. In Fig. \ref{fig0} we present the region in the two-dimensional space
$\left(  \lambda,\kappa\right)  $ where the point is real and physical acceptable.

We summarize the results of this analysis in Table \ref{tabl1}. We present the
stationary points, their physical properties as also we summarize their
stability conditions.%

\begin{table}[tbp] \centering
\caption{Stationary points and their stability for the anisotropic Chiral model}%
\begin{tabular}
[c]{ccccc}\hline\hline
\textbf{Point} & $\left(  \mathbf{\Sigma,x,y,z}\right)  $ & $\mathbf{\omega
}_{R}$ & $\mathbf{q}$ & \textbf{Stability}\\\hline
$A_{1}^{\pm}$ & $\left(  \pm1,0,0,0\right)  $ & $0$ & $2$ & Source\\
$A_{2}$ & $\left(  \frac{1}{2},0,0,0\right)  $ & $\frac{3}{4}$ & $\frac{1}{2}$
& Saddle\\
$B_{1}^{\pm}$ & $\left(  \pm\sqrt{1-x^{2}},x,0,0\right)  $ & $0$ & $2$ &
Source/Saddle\\
$B_{2}$ & $\left(  0,\frac{\lambda}{\sqrt{6}},\sqrt{1-\frac{\lambda^{2}}{6}%
},0\right)  $ & $0$ & $\frac{\lambda^{2}-2}{2}$ & Attractor Fig. \ref{fig0}\\
$B_{3}$ & $\left(  \frac{1}{2}\left(  1-\frac{3}{2\left(  1+\lambda
^{2}\right)  }\right)  ,\frac{\sqrt{6}\lambda}{2\left(  1+\lambda^{2}\right)
},\frac{\sqrt{6\left(  2+\lambda^{2}\right)  }}{2\left(  1+\lambda^{2}\right)
},0\right)  $ & $\frac{3\left(  \lambda^{2}-4\right)  }{4\left(  1+\lambda
^{2}\right)  ^{2}}$ & $\frac{\lambda^{2}-2}{2\left(  \lambda^{2}+1\right)  }$
& Attractor Fig. \ref{fig0}\\
$C_{1}$ & $\left(  0,\frac{\sqrt{6}}{2\kappa+\lambda},\sqrt{\frac{2\kappa
}{2\kappa+\lambda}},\frac{\sqrt{\lambda^{2}+2\kappa\lambda-6}}{2\kappa
+\lambda}\right)  $ & $0$ & $2-\frac{6\kappa}{2\kappa+\lambda}$ & Attractor
Fig. \ref{fig0}\\
$C_{2}$ & $\left(  2-\frac{6\kappa}{2\kappa+\lambda},\frac{\sqrt{6}}%
{2\kappa+\lambda},\frac{\sqrt{6\kappa\left(  2\kappa-\lambda\right)  }%
}{2\kappa+\lambda},\frac{\sqrt{6\kappa\lambda-3\left(  \lambda^{2}+2\right)
}}{2\kappa+\lambda}\right)  $ & $-\frac{12\kappa\left(  \kappa-\lambda\right)
}{\left(  2\kappa+\lambda\right)  ^{2}}$ & $2-\frac{6\kappa}{2\kappa+\lambda}$
& Saddle\\\hline\hline
\end{tabular}
\label{tabl1}%
\end{table}%

\section{Beyond the exponential potential}

\label{sec4}

We proceed with our analysis by considering a potential function different
from the exponential potential. In particular we assume the existence of a
cosmological constant term, such that the scalar field potential is
\begin{equation}
V\left(  \phi\right)  =V_{0}\left(  e^{-\sigma\phi}-\Lambda\right)  \text{.}
\label{ch.28}%
\end{equation}
For this potential function parameter $\lambda=\frac{V_{,\phi}}{V}$ is not a
constant, but it depends upon time variable. Indeed, $\lambda=\frac{\sigma
e^{-\sigma\phi}}{e^{-\sigma\phi}-\Lambda}$, such that $\phi=-\frac{1}{\sigma
}\ln\left(  \frac{\lambda\Lambda}{\lambda-\sigma}\right)  $. Consequently, the
derivative of $\lambda$ is different from zero, that is,
\begin{equation}
\lambda^{\prime}=\sqrt{6}x\lambda\left(  \sigma-\lambda\right)  \text{.}
\label{ch.29}%
\end{equation}

Therefore, for a non-exponential scalar field potential, the gravitational
field equations have one extra dimension, i.e. equation (\ref{ch.29}). Because
of the existence of equation (\ref{ch.29}) new stationary points follow, but
the stability properties of the previous points may change.

As above, we summarize the stationary points $P=\left(  \Sigma\left(
P\right)  ,x\left(  P\right)  ,y\left(  P\right)  ,z\left(  P\right)
,\lambda\left(  P\right)  \right)  $ on three families of points, families
$\bar{A}$ ,~$\bar{B}$ and $\bar{C}$. The stationary points are categorized
according to the contribution of the scalar field in the cosmological fluid as above.

\subsection{Stationary points of the family $\bar{A}$}

The family $\bar{A}$ is defined by the stationary points
\begin{equation}
\bar{A}_{1}^{\pm}=\left(  \pm1,0,0,0,\lambda\right)  ~,~\bar{A}_{2}=\left(
\frac{1}{2},0,0,0,\lambda\right)  ~,~\lambda~\text{\ arbitrary.}%
\end{equation}

The physical properties of the asymptotic solutions at the stationary points
are similar to those of the exponential potential. The eigenvalues for the
five-dimensional linearized system are derived to be $e_{1}\left(  \bar{A}%
_{1}^{+}\right)  =6$~,$~e_{2}\left(  \bar{A}_{1}^{+}\right)  =3$%
,$~e_{3}\left(  \bar{A}_{1}^{+}\right)  =0$~,~$e_{4}\left(  \bar{A}_{1}%
^{+}\right)  =0$,~$e_{5}\left(  \bar{A}_{1}^{+}\right)  =0~$; $e_{1}\left(
\bar{A}_{1}^{-}\right)  =3$~,~$e_{2}\left(  \bar{A}_{1}^{-}\right)  =2$,
$e_{3}\left(  \bar{A}_{1}^{-}\right)  =0$,~ $e_{4}\left(  \bar{A}_{1}%
^{-}\right)  =0$~,~$e_{5}\left(  \bar{A}_{1}^{-}\right)  =0$ ; $e_{1}\left(
\bar{A}_{2}\right)  =\frac{3}{2}~,~e_{2}\left(  \bar{A}_{2}\right)  =-\frac
{3}{2}$, $e_{3}\left(  \bar{A}_{2}\right)  =-\frac{3}{2}$ ,$~e_{4}\left(
\bar{A}_{2}\right)  =-\frac{3}{2}$ ,$~~e_{5}\left(  \bar{A}_{2}\right)  =0$.
We observe that the stability properties do not change for the stationary
points. Hence, the points $\bar{A}_{1}^{\pm}$ are sources, while the point
$\bar{A}_{2}$ is a saddle point.

\subsection{Stationary points of family $\bar{B}$}

Family $\bar{B}$ is composed of the following points%

\[
\bar{B}_{1}^{\pm}=\left(  \pm\sqrt{1-x^{2}},x,0,0,\sigma\right)  ~,~\bar
{B}_{2}=\left(  0,\frac{\sigma}{\sqrt{6}},\sqrt{1-\frac{\sigma^{2}}{6}%
},0,\sigma\right)  ,
\]

\[
\bar{B}_{3}=\left(  \frac{1}{2}\left(  1-\frac{3}{2\left(  1+\sigma
^{2}\right)  }\right)  ,\frac{\sqrt{6}\sigma}{2\left(  1+\sigma^{2}\right)
},\frac{\sqrt{6\left(  2+\sigma^{2}\right)  }}{2\left(  1+\sigma^{2}\right)
},0,\sigma\right)  ~,
\]%
\[
\bar{B}_{4}^{\pm}=\left(  \pm\sqrt{1-x^{2}},x,0,0,0\right)  ~,~\bar{B}%
_{5}=\left(  -1,0,\sqrt{3},0,0\right)  ~,~\bar{B}_{6}=\left(
0,0,1,0,0\right)  .
\]

The stationary points, $\bar{B}_{1}^{\pm}$,~$\bar{B}_{2}$ and $\bar{B}_{3}$,
have the same physical properties with the corresponding points of family $B$.
Thus we study only their stability properties.

The eigenvalues for the linearized system around $\bar{B}_{1}^{\pm}$ are
$e_{1}\left(  \bar{B}_{1}^{\pm}\right)  =2\left(  2\mp\sqrt{1-x^{2}}\right)
$~,$~e_{2}\left(  \bar{B}_{1}^{\pm}\right)  =\sqrt{6}x\kappa$,$~e_{3}\left(
\bar{B}_{1}^{\pm}\right)  =-\sqrt{6}x\sigma$~,~$e_{4}\left(  \bar{B}_{1}^{\pm
}\right)  =\frac{1}{2}\left(  6-\sqrt{6}x\sigma\right)  $,~$e_{5}\left(
\bar{B}_{1}^{\pm}\right)  =0$. Therefore, the family of points $\bar{B}%
_{1}^{\pm}$ are saddle points. Moreover, the eigenvalues around $\bar{B}_{2}$
are $e_{1}\left(  \bar{B}_{2}\right)  =\frac{\sigma^{2}+2\kappa\sigma-6}{2}%
$~,$~e_{2}\left(  \bar{B}_{2}\right)  =\frac{\sigma^{2}-6}{2}$,$~e_{3}\left(
\bar{B}_{2}\right)  =-\sqrt{6}x\sigma$~,~$e_{4}\left(  \bar{B}_{2}\right)
=\sigma^{2}-2$,~$e_{5}\left(  \bar{B}_{2}\right)  =-\sigma^{2}$, that is,
point $\bar{B}_{2}$ has similar stability properties with point $B_{2}$, as
presented in Fig. \ref{fig0}. Moreover, for point $\bar{B}_{3}$ we calculate
the eigenvalues $e_{1}\left(  \bar{B}_{3}\right)  =-\frac{3\left(
2-2\kappa\sigma+\sigma^{2}\right)  }{2\left(  1+\sigma^{2}\right)  }$ ,
$e_{2}\left(  \bar{B}_{3}\right)  =-\frac{3\left(  2+3\sigma^{2}+\sigma
^{4}\right)  }{2\left(  1+\sigma^{2}\right)  }$ ,~$e_{3}\left(  \bar{B}%
_{3}\right)  =-\frac{3\left(  2+3\sigma^{2}+\sigma^{4}+\sqrt{\left(
1+\sigma\right)  ^{2}\left(  2+\sigma^{2}\right)  \left(  7\sigma
^{2}-18\right)  }\right)  }{4\left(  1+\sigma^{2}\right)  ^{2}}$,
$e_{4}\left(  \bar{B}_{3}\right)  =-\frac{3\left(  2+3\sigma^{2}+\sigma^{4}%
\pm\sqrt{\left(  1+\sigma\right)  ^{2}\left(  2+\sigma^{2}\right)  \left(
7\sigma^{2}-18\right)  }\right)  }{4\left(  1+\sigma^{2}\right)  ^{2}}$ and
$e_{5}\left(  B_{3}\right)  =-\frac{3\sigma^{2}}{1+\sigma^{2}}$. The
eigenvalue $e_{5}\left(  \bar{B}_{3}\right)  $ is always negative, while the
rest are similar to those of point $B_{3}$. Thus, point $\bar{B}_{3}$ is an
attractor in the region presented in Fig. \ref{fig0}, where $\lambda=\sigma$.

The stationary points $\bar{B}_{4}^{\pm}$, $\bar{B}_{5}$,~$\bar{B}_{6}$
correspond to asymptotic solutions for which the scalar field potential plays
the role of the cosmological constant, that is $V_{,\phi}=0$, and $V\left(
\phi\right)  =const.$ Points $\bar{B}_{4}^{\pm}$ have the same physical
properties with $\bar{B}_{1}^{\pm}$, while the eigenvalues of the linearized
system are the same, for $\sigma=0$.

The point $\bar{B}_{5}$ describes an anisotropic inflationary exact solution
in a Kantowski-Sachs spacetime, $\omega_{R}\left(  \bar{B}_{5}\right)  =-3$,
$q\left(  \bar{B}_{5}\right)  =-1$. The eigenvalues are $e_{1}\left(  \bar
{B}_{5}\right)  =-6$, $e_{2}\left(  \bar{B}_{5}\right)  =-3$, $e_{3}\left(
\bar{B}_{5}\right)  =-3$,~$e_{4}\left(  \bar{B}_{5}\right)  =3$, $e_{5}\left(
\bar{B}_{5}\right)  =0$, which means that $\bar{B}_{5}$ is a saddle point.

\begin{figure}[ptb]
\centering\includegraphics[width=1\textwidth]{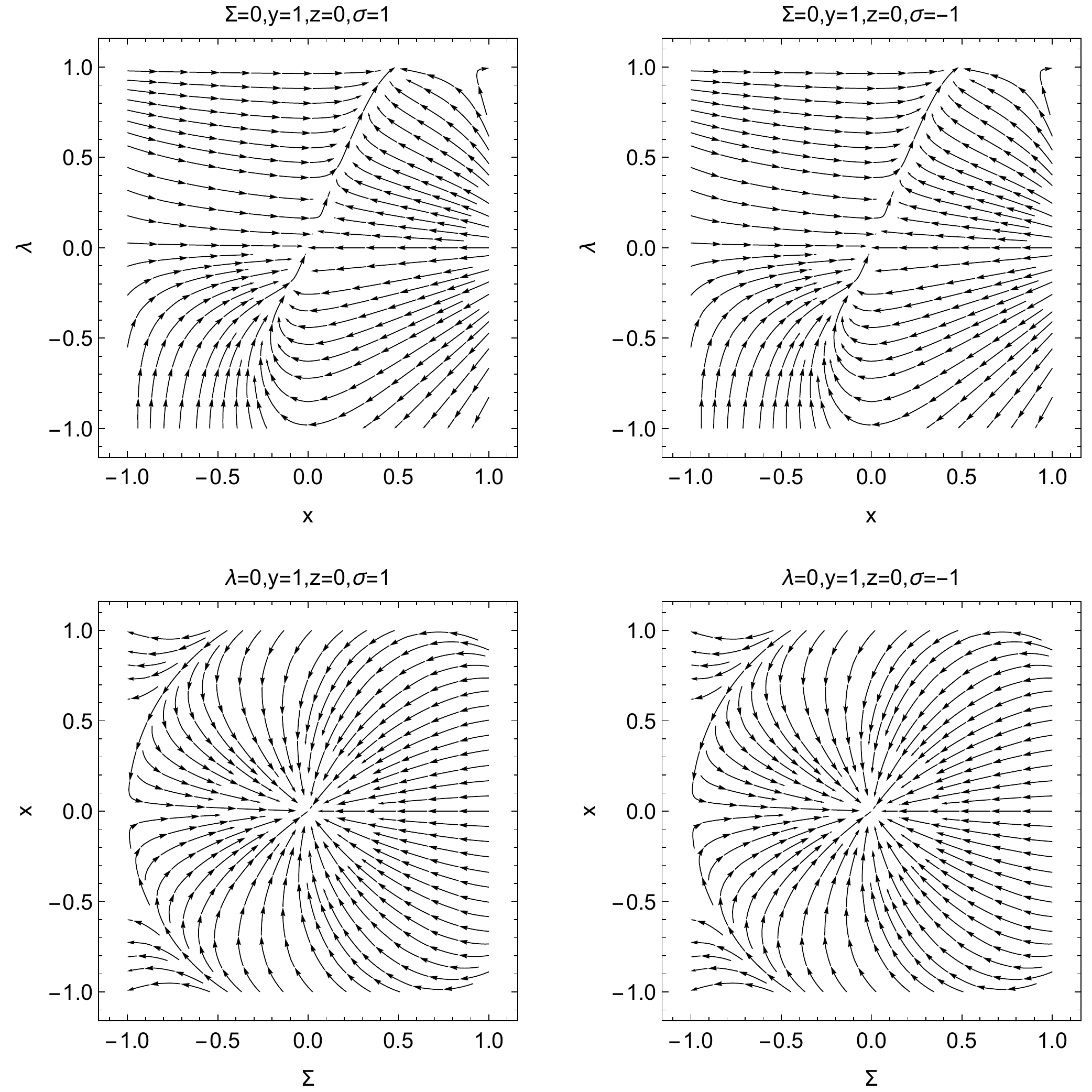}\caption{Phase-space
portraits for the dynamical system around the stationary point $\bar{B}_{6}$.
Left figures are for $\sigma=+1$, while right figures are for $\sigma=-1$.
Figures of the first row are in the plane $\left\{  x,\lambda\right\}  $,
where we observe that $\bar{B}_{6}$ is a saddle point. However, because the
dynamical system at the stationary point has four eigenvalues with negative
real parts, there is a stable submanifold when $\lambda=0$, as it can be seen
from the figures of the second row. }%
\label{fig02}%
\end{figure}

Finally, points $\bar{B}_{6}$ describe the de Sitter solution in a spatially
flat FLRW spacetime, $\omega_{R}\left(  \bar{B}_{6}\right)  =0$, $q\left(
\bar{B}_{6}\right)  =-1$. The eigenvalues of the linearized system are
$e_{1}\left(  \bar{B}_{6}\right)  =-3$, $e_{2}\left(  \bar{B}_{6}\right)
=-3$, $e_{3}\left(  \bar{B}_{6}\right)  =-3$,~$e_{4}\left(  \bar{B}%
_{6}\right)  =-2$, $e_{5}\left(  \bar{B}_{6}\right)  =0$. In Fig. \ref{fig02}
we discuss the stability properties for the stationary point $\bar{B}_{6}$. We
observe that the point is in general a saddle space, but it has a stable
manifold in the subspace $\left\{  \Sigma,x,y,z\right\}  $ for $\lambda=0$.
The exact form for the stable manifold can be derived with the application of
the center manifold theorem. We select to omit such presentation because it
does not contribute in the physical discussion for the anisotropic model.

\subsection{Stationary points of family $\bar{C}$}

The stationary points with $xz\neq0$ are
\[
\bar{C}_{1}=\left(  0,\frac{\sqrt{6}}{2\kappa+\lambda},\sqrt{\frac{2\kappa
}{2\kappa+\lambda}},\frac{\sqrt{\lambda^{2}+2\kappa\lambda-6}}{2\kappa
+\lambda},\sigma\right)  ~
\]
and
\[
\bar{C}_{2}=\left(  2-\frac{6\kappa}{2\kappa+\lambda},\frac{\sqrt{6}}%
{2\kappa+\lambda},\frac{\sqrt{6\kappa\left(  2\kappa-\lambda\right)  }%
}{2\kappa+\lambda},\frac{\sqrt{6\kappa\lambda-3\left(  \lambda^{2}+2\right)
}}{2\kappa+\lambda},\sigma\right)  ~\text{.}%
\]

Thus the physical properties of the solutions are similar those of points
$C_{1}$ and $C_{2}^{\pm}$, respectively. Moreover, the stability properties
are the same as above. Indeed, point $C_{1}$ is an attractor as presented in
Fig. \ref{fig0} while point $C_{2}$ is always a saddle point.

Finally, there are no (real valued) stationary points for $\lambda=0$.

We conclude that the consideration of a different potential function different
from the exponential, provides new stationary points only in family $B$, that
is, of the quintessence case, in which the second scalar field does not
contribute in the cosmological fluid, $z=0$. The latter can be easily seen
and, if we consider an arbitrary potential function~$V\left(  \phi\right)  $,
where no new stationary points in the family $C$ follow.

Finally, the new physical solutions are anisotropic inflationary solutions in
a Kantowski-Sachs spacetime with cosmological constant and the de Sitter universe.

\section{Conclusions}

\label{sec5}

We performed a detailed analysis on the dynamics for the Chiral cosmological
theory in a anisotropic background space. The Chiral model belongs to the
multi scalar field theories, in which the energy-momentum tensor of the theory
is consisted by two interacting scalar fields minimally coupled to the
gravity. The two scalar fields interact in the kinetic part, such that the
scalar fields to lie on the hyperbolic plane.

For this model, and for the generic LRS background space which describes the
Bianchi I, the Bianchi\ III and the Kantowski-Sachs spacetimes we wrote the
field equations by using dimensionless variables in the $H$-normalization
approach. Because of the large number of the dependent variables, we selected
to work on the $H$-normalization instead of other dimensionless variables. We
determined the stationary points of the field equations and we investigated
their stability properties. Every stationary point corresponds to a specific
exact solution for the field equations which describe a specific asymptotic
behaviour during the cosmological evolution.

The stationary points have been categorized in three families. Points of
family $A$ describe the limit of General Relativity without matter source,
points of family $B$ correspond to the stationary points with a quintessence
matter source, while points of the third family, namely family $C$, describe
exact solutions where the two fields contributes. The points of the third
family are of special interests because they can describe isotropic and
anisotropic inflationary solutions with two scalar fields. In particular we
recovered the isotropic inflationary model known hyperbolic inflation
\cite{vr91}, while the anisotropic inflationary solution can be sees as the
analogue of hyperbolic inflation. Because the anisotropic hyperbolic
inflationary solution is always unstable, point $C_{2}$ is a saddle point, we
can say that the anisotropic inflationary solution played role in the very
early universe such is in the beginning of the inflation.

This work contributes on the anisotropic inflationary models. In a future
study we plan to investigate further the effects of the Chiral model in
anisotropic background spaces.

\begin{acknowledgments}
The research of AG was funded by Agencia Nacional de Investigaci\'{o}n y
Desarrollo - ANID through the program FONDECYT Regular grant no. 1200293. The
research of AP and GL was funded by Agencia Nacional de Investigaci\'{o}n y
Desarrollo - ANID through the program FONDECYT Iniciaci\'{o}n grant no.
11180126. Additionally, GL was funded by Vicerrector\'{\i}a de
Investigaci\'{o}n y Desarrollo Tecnol\'{o}gico at Universidad Catolica del
Norte. This work is based on research supported in part by the National
Research Foundation of South Africa (Grant Numbers 131604).
\end{acknowledgments}


\begin{thebibliography}{99}                                                                                               %


\bibitem {mf0}A.A. Coley and R.J. van den Hoogen, Phys. Rev. D 62, 023517 (2000)

\bibitem {mf1}H. Abedi and A.M. Abbasi, JCAP 07, 049 (2017)

\bibitem {mf2}M. Sasaki and T. Tanaka, Prog. Ther. Phys. 99, 763 (1998)

\bibitem {mf3}C. van de Bruck, A.J. Christopherson and M. Robinson, Phys. Rev.
D 91, 123503 (2015)

\bibitem {mf4}F. Galli and A.S. Koshelev, Theor. Math. Phys. 164, 1169 (2010)

\bibitem {mf5}C. van de Bruck and M. Robinson, JCAP 08, 024 (2014)

\bibitem {mf6}L.P. Chimento, A.E. Cossarini and N.A. Zuccala, Class. Quantum
Grav. 15, 57 (1998)

\bibitem {mf7}C.R. Fadragas, G. Leon and E.N. Saridakis, Class. Quantum Grav.
31, 075018 (2014)

\bibitem {mf8}J. Socorro, S. P\'{e}rez-Pay\'{a}n, R. Hern\'{a}ndez, A. Class.
Quantum Grav. 38, 135027 (2021)

\bibitem {mf9}J. Socorro and O.E. Nu\~{n}ez, Eur. Phys. J. Plus, 132, 168 (2017)

\bibitem {mf10}P.~Christodoulidis, Eur. Phys. J. C 81, 471 (2021)

\bibitem {mf11}M.\ Rainar and A. Zhuk, Phys.\ Rev. D 54, 6186 (1996)

\bibitem {mf12}L.R. D\'{\i}az-Barr\'{o}n, A. Espinoza-Garc\'{\i}a and J.
Socorro, to appear in Int. J. Mod. Phys. D (2021) 10.1142/S0218271821500802

\bibitem {guth}A. Guth, Phys. Rev. D 23, 347 (1981)

\bibitem {g1}K.Y. Choi, S.A. Kim and B. Kyae, Nucl. Phys. B 861, 271 (2021)

\bibitem {g2}D.H. Lyth, JCAP 0511, 006 (2005)

\bibitem {g3}D.\ Wands, Lect. Notes Phys. 738, 275 (2008)

\bibitem {g4}D.I. Kaiser, E.A. Mazenc and E.I. Sfakianakis, Phys. Rev. D 87,
064004 (2013)

\bibitem {g5}D. Langlois and S. Renaux-Peterl, JCAP 0804, 017 (2008)

\bibitem {qq2}Y.-F. Cai, E.N. Saridakis, M.R. Setare and J.-Q. Xia,
Phys.\ Rept. 493, 1 (2010)

\bibitem {ancqg}A. Paliathanasis, Class. Quantum Grav. 37, 195014 (2020)

\bibitem {atr6}S.V. Chervon, Quantum Matter 2, 71 (2013)

\bibitem {vr91}A.~R.~Brown, Phys. Rev. Lett. 121, 251601 (2018)

\bibitem {vr92}S.~Mizuno and S.~Mukohyama, Phys. Rev. D 96, 103533 (2017)

\bibitem {s1}V. Aragam, S. Paban and R. Rosati, JHEP 9, 2021 (2021)

\bibitem {ts1}A. Paliathanasis and M.\ Tsamparlis, Phys.\ Rev. D 90, 043529 (2014)

\bibitem {ts2}P. Christodoulidis, D. Roest and E.I. Sfakianakis, JCAP 1911,
002 (2019)

\bibitem {ts3}S.V. Chernov and N.A. Koshelev, Grav. Cosmol. 9, 196 (2003)

\bibitem {ts4}R.A. Abbyazov and S.V. Chernov, Mod. Phys. Lett. A 28, 1350024 (2013)

\bibitem {ts5}A.\ Paliathanasis, G.\ Leon and S. Pan, Gen. Rel. Gravit. 51,
106 (2019)

\bibitem {ts6}M. Cicoli, G. Dibitetto and P.G. Pedro, Phys.\ Rev. D 101,
103524 (2020)

\bibitem {ts7}M. Cicoli, G. Dibitetto and P.G. Pedro, JHEP 10, 35 (2020)

\bibitem {ts8}S. Mizuno, S. Mukohyama, S. Pi and Y.L Zhang, JCAP 09, 072 (2019)

\bibitem {ts9}N. Dimakis, A. Paliathanasis, P.A.\ Terzis and T.
Christodoulakis, EPJC 79, 618 (2019)

\bibitem {ts10}V.R. Ivanov and S.Yu Vernov, Integrable modified gravity
cosmological models with an additional scalar field, (2021) [arXiv:2108.10276]

\bibitem {Mis69}\ C.W. Misner, Astroph. J. 151, 431 (1968)

\bibitem {jacobs2}K.C. Jacobs, Astrophys J. 153, 661 (1968)\ 

\bibitem {collins}C.B Collins and S.W. Hawking, Astroph. J. 180, 317 (1973)

\bibitem {wald}R.M. Wald, Phys Rev. 28, 2118 (1983)

\bibitem {nilson}U. Nilsson and C. Uggla, Clas. Quantum Grav. 13, 1601 (1996)

\bibitem {kas2}S.M.M. Rasouli, M. Farhoudi and H.R. Sepangi, Class. Quantum
Grav. 28, 155004 (2011)

\bibitem {kas3}X.O. Camanho, N. Dadhich and A. Molina, Class. Quantum Grav.
32, 175016 (2015)

\bibitem {kas4}P. Halpern, Phys. Rev. D 63, 024009 (2001)

\bibitem {kas5}L.E. Mendes and A.B. Henriques, Phys.\ Lett. B 254, 44 (1991)

\bibitem {kas6}M. Kar\v{c}iauskas, Mod. Phys. Lett. A, 31, 1640002 (2016)

\bibitem {kas7}A. Talebian, A. Nassiri-Rad and H. Firouzjahi, Phys. Rev. D
101, 023524 (2020)

\bibitem {kas8}A.A. Abolhasani, R. Emami and H. Firouzjahi, JCAP 05, 016 (2014)

\bibitem {kas9}R.K. Tiwari, A. Beesham, S. Mishra and V. Dubey, Universe 7,
226 (2021)

\bibitem {kas10}G. Leon, A. Paliathanasis and N. Dimakis, EPJC 80, 1149 (2020)

\bibitem {aco}A.A. Coley, Dynamical Systems and Cosmology, Astrophysics and
Space Science Library 291, Springer Netherlands, Amsterdam, (2003)

\bibitem {sc1}J. Wainwright and G.F.R. Ellis, Dynamical Systems in Cosmology,
Cambridge University Press, Cambridge (1997)

\bibitem {sc2}E.J Copeland, A.R. Liddle and D. Wands, Phys. Rev.\ D 57, 4686 (1998)

\bibitem {sc3}R. Lazkoz, G. Leon and I. Quiros, Phys. Lett. B 649, 103 (2007)

\bibitem {sc4}G. Leon, J. Saavedra and E.N. Saridakis, Class. Quantum Grav.
30, 135001 (2013)

\bibitem {sc5}A. Paliathanasis, Phys. Rev. D 101, 064008 (2020)

\bibitem {sc6}G. Leon and E.N. Saridakis, JCAP 03, 025 (2013)

\bibitem {sc7}G. Papagiannopoulos, S. Basilakos, A. Paliathanasis, S. Pan and
P. Stavrinos, EPJC 80, 816 (2020)

\bibitem {sc8}C.R. Fadragas, G. Leon and E.N. Saridakis, Class. Quantum Grav.
31, 075018 (2014)

\bibitem {sc9}N. Goheer, J.A. Leach and P.K.S. Dunsby, Class. Quantum Grav.
24, 5689 (2007)

\bibitem {sc10}J.D. Barrow and A. Paliathanasis, EPJC 78, 767 (2018)

\bibitem {sc11}G. Leon and A.A.\ Roque, JCAP 05, 032 (2014)

\bibitem {sc12}J. Latta, G.\ Leon and A. Paliathanasis, JCAP 11, 051 (2016)

\bibitem {sc13}D. Shogin and S. Hervik, Class. Quantum Grav. 32, 055008 (2015)

\bibitem {sc14}R.J.\ van den Hoogen, A.A. Coley, B. Alhulaimi, S. Mohandas, E.
Knighton and S. O'Neil, JCAP 11, 017 (2018)

\bibitem {anuni1}A. Paliathanasis, Universe 7, 323 (2021)
\end{thebibliography}
\end{document}